 \documentclass[preprint,aps,prl,showpacs,preprintnumbers,amsmath,amssymb,superscriptaddress]{revtex4-1}
\usepackage{graphicx,textcomp}  
\usepackage{bm}        
\usepackage{amssymb,multirow}   
\usepackage[outdir=./]{epstopdf}
 \usepackage[usenames,dvipsnames]{xcolor}
 \usepackage[hyperfootnotes=false]{hyperref}
 \usepackage[hyperfootnotes=false]{hyperref}
 \hypersetup{  colorlinks=true,  citecolor=Violet,  linkcolor=Blue,  filecolor= green,  urlcolor=Blue}
\usepackage{dcolumn}
\usepackage{bm}
\usepackage{multirow}
\usepackage{subfigure}
\usepackage{xcolor}
\usepackage[normalem]{ulem}
\usepackage{soul}

\begin{document}

\title{Microscopic Origin Of Room Temperature Ferromagnetism in a Double Perovskite Sr$_2$FeReO$_6$: a first principle and model Hamiltonian study}

\author{Shishir K. Pandey}
\affiliation{Department of Condensed Matter Physics and Material Science,  S. N. Bose National Center for Basic Sciences,Kolkata-700106}

\author{Ashis K. Nandy}
\affiliation{Department of Condensed Matter Physics and Material Science,  S. N. Bose National Center for Basic Sciences,Kolkata-700106}
\affiliation{School of Physical Sciences, National Institute of Science Education and Research, HBNI, Jatni-752050, India}

 \author{Poonam Kumari}
 \affiliation{Department of Condensed Matter Physics and Material Science,  S. N. Bose National Center for Basic Sciences,Kolkata-700106}
 
\author{Priya Mahadevan}
\email{priya.mahadevan@gmail.com}
\affiliation{Department of Condensed Matter Physics and Material Science,  S. N. Bose National Center for Basic Sciences,Kolkata-700106}

\begin{abstract}
The puzzling observation of room temperature ferromagnetism in double perovskites (A$_2$BB$'$O$_6$), despite having the magnetic lattice 
of B-ions diluted by non-magnetic B$'$-ions, have been examined for Sr$_2$FeReO$_6$. {\it Ab-initio} spin spiral electronic structure calculations along various high symmetry directions in reciprocal space are used to determine the exchange interactions entering an extended Heisenberg model, which is then solved classically 
using Monte Carlo simulations to determine the ferromagnetic transition temperature T$_c$. We find that one must consider onsite Coulomb interactions 
at the nonmagnetic Re sites ($U$) in order to obtain a T$_c$ close to the experimental value. Analysis of the $ab$-$initio$ electronic structure as well as an 
appropriate model Hamiltonian trace the origin of enhancement in T$_c$ with $U$ to the enhanced exchange splitting that is introduced at these sites. This in turn destabilizes the antiferromagnetic exchange channels, thereby enhancing the T$_c$. The role of occupancy  at the non-magnetic sites is examined by contrasting with the case of Sr$_2$FeMoO$_6$.
\end{abstract}

\maketitle

\section{Introduction}
Double perovskite (DP) oxides have the general formula A$_2$BB$'$O$_6$ with A being the rare earth or alkaline-earth ion, B and B$'$ are the transition metal atoms separated 
by oxygen atoms in the lattice. In general, a DP oxide consists of two sub lattices of perovskite ABO$_3$ and AB$'$O$_3$ units in a way that the B and B$'$ sites
are arranged alternately 
in a three dimensional array. B (B$'$) transition metal atoms reside in an octahedra formed by six oxygen atoms. The finding of ferromagnetic behavior in the Re based DP 
oxides A$_2$FeReO$_6$ (where A= Ba, Sr, Ca) above room temperature (with ferromagnetic transition temperatures 316, 401, 538 K respectively) \cite{ch_sfro_ref12} stimulated  
research in 
this new ferromagnetic class of DP oxides. Another similar example of a high T$_c$ ferromagnetic DP oxide is Sr$_2$FeMoO$_6$ (SFMO) with a transition temperature of 410 K 
\cite{ch_sfro_ref14,ch_sfro_ref14a}. 
In Sr$_2$FeReO$_6$ (SFRO), the separation between the closest magnetic Fe atoms is $\sqrt{2}$ times larger than that found in a Fe-based perovskite oxide. This large 
separation of Fe atoms arises because of the dilution of the magnetic lattice with non-magnetic Re atoms. The naive expectation was that it should result in weaker magnetic interactions and hence a lower 
magnetic ordering temperature. Surprisingly,  the Curie temperature for this system is 401 K. In contrast, considering the 
case of the hole doped manganites \cite{manganites_rev}, where one has smaller separations between the magnetic atoms, one finds  a 
lower magnetic ordering temperature. As the ferromagnetism has been conventionally understood within a double exchange model, 
the presence of nonmagnetic atoms like Re/Mo present in a double perovskite crystal 
requires a different mechanism to explain the high ferromagnetic ordering temperature in these 
compounds. 

A qualitative model to understand the high magnetic ordering temperature was proposed by Sarma {\it et al.} \cite{ch_sfro_ref16} after 
analyzing the $ab$ $initio$ electronic structure of SFMO. They concluded that the hopping interactions between first neighbor Fe and Mo atoms in the lattice led to an 
exchange splitting being induced on the Mo sites which is opposite in direction to that at the Fe site. This led to an antiferromagnetic coupling between the Fe and Mo sites 
and an effective ferromagnetic coupling between the Fe sites. This results in the high magnetic ordering temperatures of these materials. 

Electronic structure calculations reveal that the majority Fe-up spin $t_{2g}$ and $e_g$ states are $\sim$ 3.75 eV and 2.25 eV 
below the fermi energy respectively, while the minority spin 
states have highly hybridized Fe-Re/Mo character. This idea led to the use a Kondo-like model to describe the magnetism in these materials. 
The majority spin Fe states are approximated by a spin within this model which interacts with the itinerant conduction electrons through an
antiferromagnetic coupling. The itinerant conduction electron can delocalize by hopping from the $t_{2g}$ levels 
on the nonmagnetic transition metal atom to those on Fe. While this model was able to successfully explain the high magnetic ordering temperature in Sr$_2$FeMoO$_6$ which had just 
one electron on the nonmagnetic atom, it failed to explain the high magnetic ordering temperature when one had more than one electron on the nonmagnetic 
atom as in Sr$_2$FeReO$_6$ \cite{ch_sfro_ref17,ch_sfro_ref18,ch_sfro_ref19,ch_sfro_ref20}. 

Brey and coworkers \cite{ch_sfro_ref21} carried out a mean field study considering a  $t_{2g}$ only model. 
The suppression of Tc  was recovered qualitatively by introducing  electron-electron interactions at the nonmagnetic site within the model. This helped in obtaining the experimental trend of increase in T$_c$ with increase of the number of electrons at the nonmagnetic site. We reexamine the changes in the electronic structure at the nonmagnetic site, examining how it evolves with $U$ and how this affects the stability of the ferromagnetic state with respect to the competing antiferromagnetic state, in addition to the modifications in the exchange interaction 
strengths. 

In the context, the issue of high temperature magnetism in Sr2FeReO6  has been studied within ab initio electronic structure as well as a model Hamiltonian approach. The $ab$ $initio$ band structure calculations were performed 
 and electron correlation effects on Fe as well as on Re sites were introduced within the GGA+U technique. Considering a $U$ of 2.5 eV on the Fe site, the ground state is 
 found to be ferromagnetic. The magnetic exchange interaction strengths ($J$'s) between the magnetic (Fe) atoms were calculated using an inverse Fourier transform method. 
 Using a classical Monte Carlo (MC) simulation 
 technique employing an extended Heisenberg model, the T$_c$ of the system was estimated.  In the first case, when there was no $U$ on the Re site, we found that the nearest neighbor 
 exchange interaction $J_1$ is ferromagnetic while the second neighbor exchange interaction $J_2$ is antiferromagnetic contrary to what is believed in literature \cite{ch_sfro_ref22}. 
 $J_3$ and $J_4$ are negligibly small when compared to $J_1$, $J_2$. With these exchange interaction strengths, the estimated T$_c$ within the Monte Carlo simulation study 
 was much lower ($\sim$ 285 K) than the experimental value of 401 K. Taking a clue from the work by Brey {\it et al.}~\cite{ch_sfro_ref21}, we then considered a $U$ of 1 eV on the Re-$d$ states and again 
 calculated the exchange interaction strengths. We found that the $J_1$ got enhanced by a factor 
 of $\sim$ 2. Most surprisingly, $J_2$ changes its sign from being antiferromagnetic in the no $U$ case to ferromagnetic in the present case when a $U$ of 1 eV is considered at the Re 
 site. Using a $U$(Fe) = 2.5 eV and $U$(Re) = 1.0 eV, we find the magnetic ordering temperature to be 395 K which is reasonably close to the experimentally observed value. 
 Within $ab$ $initio$ electronic structure calculations, the ferromagnetic ground state is found to be half-metallic. The Re moments 
 found in this case are induced in nature and are 
 antiparallel to Fe moments. With an application of $U$ on the Re site, we found no substantial effect in the density of states for the ferromagnetic configuration which could 
 explain the gain in stability in this case. However, examining the density of states in a competing antiferromagnetic configuration we found that the $U$(Re) played the 
 role on inducing an exchange splitting of the Re-$d$ states. Thus by destabilizing the antiferromagnetic state, one found that one could increase the ferromagnetic 
 stability.  In order to test this hypothesis we set up a multiband Hubbard Hamiltonian to describe the system. Introducing an exchange splitting ($J_h$) on Re  was 
found to enhance the stability of the ferromagnetic state.

\section{Methodology}
 
 The electronic and magnetic properties of SFRO have been calculated using a plane wave pseudopotential implementation of density functional theory using PAW potentials\cite{ch_sfro_ref26,ch_sfro_ref27} implemented in  the Vienna ab-initio simulation package (VASP). The generalised
 gradient approximation (GGA) \cite{ch_sfro_ref28,ch_sfro_ref29} has been used for the exchange correlation functional. The effects of electron-electron correlation on the transitional metal $d$ electrons were considered within the GGA+U scheme using the Dudarev {\it et al.} formalism \cite{ch_sfro_ref30} for the plus U implementation. In our calculations we have used a $U$ of 2.5 eV on the Fe-$d$ states while a $U$ of 0, 1 and 2~eV was considered on the Re-$d$ states. A k-mesh of 8 $\times$ 8 $\times$ 8 was used for the integrations in momentum space, while a cutoff energy of 500 eV was used for the plane wave included in the basis for the calculations. The inter atomic exchange interactions 
strengths J's were determined using the frozen magnon approach\cite{ch_sfro_ref31,ch_sfro_ref32,ch_sfro_ref33,ch_sfro_ref34}. In this approach the total energy corresponding to 
 each spin spiral state associated with a spin spiral vector {$\vec {q}$} was calculated within a 
 noncollinear spin implementation in VASP\cite{ch_sfro_ref35}. 
Each spin at the Fe site is treated classically and treated as a vector where {$\vec{s_i}$} is a unit vector along the direction 
of the spin at the site labelled by $i$. Hence, 
 \begin{equation}
 \vec{s_i} = ({\sin\theta \cos(\vec{q} \cdot \vec{R_i}), \sin\theta \sin(\vec{q} \cdot \vec{R_i}), \cos\theta})
 \end{equation}
where $\vec{R_i}$ is the position of the atom  and $\theta$ is the polar angle made by the spin with the z-axis. Assuming the spins at the 
 Fe sites to be localized, we mapped the magnetic energies calculated for various {$\vec{q}$} onto an extended Heisenberg model as, 
 \begin{equation}
 H = -\frac{1}{2}\sum_{ij}J_{ij} \vec{s_i}\cdot \vec{s_j}
 \end{equation}
In the above equation, $s_{i}$, $s_{j}$ are unit vectors in the direction of the spins at sites i and j while $J_{ij}$  
are the exchange interaction strengths between the Fe spins at site $i$ and $j$. The magnitude of spins are absorbed into the $J$. It is clear 
 that in this model, positive and negative $J$ value imply ferromagnetic and antiferromagnetic exchange interactions respectively. Hence, corresponding to each $\vec{q}$ value, 
 one has the total energy E($\vec{q}$) in terms of the J($\vec{q}$) which is related to the exchange interactions via Fourier transformation as, 
 \begin{equation}
 J(\vec{q}) = \sum_{i\ne 0} J_{0i} e^{i\vec{q}\cdot \vec{R_{0i}}}
 \end{equation}

For simplicity, in our calculations, we restricted the spins to lie in the xy plane 
After performing an inverse Fourier transform, we obtain the real space exchange parameters as, 
 \begin{equation}
 J_{0i} = \frac{1}{N} \sum_{\vec{q}} J(\vec{q}) e^{-i\vec{q}\cdot \vec{R_{0i}}}
 \end{equation}
 Thus the calculated spin wave spectra along various various high symmetry directions with varying $\vec{q}$  has been determined and the spectrum   was fitted using a least squared error minimization procedure to estimate the exchange parameters with the best fit. The Boltzmann weight factor $e^{-\beta \Delta E}$ was considered while estimating the exchange 
 parameters to take care of the fact that statistical weight corresponding to large angles between the spins are low in the frozen magnon states. Here $\beta = 1/k_B T$ where 
 $T$ is the ferromagnetic transition temperature of 401 K for SFRO and  $\Delta E = E(\vec{q})-E(0)$ ~\cite{thesis_ashis}.
This analysis was done for both the case when there was no $U$ on the Re sites 
 and also with a $U$(Re) = 1 eV while the $U$ on Fe sites was kept fixed at 2.5 eV.

With the help of these exchange interactions, we have calculated the ferromagnetic transition temperature (T$_c$) using a classical Monte Carlo (MC)\cite{ch_sfro_ref36} method. 
In this technique, we first set up the magnetic lattice. In SFRO the Fe atoms form a face centered lattice. A system size consisting of 16 $\times$ 16 $\times$ 16 spins is considered. The extracted $J$'s from the procedure discussed above are used as inputs for an extended Heisenberg model. 
During a single MC run, random orientations ($\vec{s_i}$) for a spin at $i$-th site were considered and the energy of the system was calculated.
Then a new random orientation of spin ($\vec{s'_i}$) was created and the system energy was again calculated. If the energy associated with this new random orientation ($\vec{s'_i}$) was lower than that of ($\vec{s_i}$), then ($\vec{s'_i}$) was accepted. Else, Metropolis algorithm\cite{ch_sfro_ref37} was employed to decide whether to accept or reject ($\vec{e'_i}$).  
This procedure was performed over all the lattice sites $i$ is one Monte Carlo step. The magnetization M(T) per spin was calculated only after bringing the system into thermal equilibrium. Total 5 $\times$ 10$^5$ Monte Carlo steps were used in these calculations and results were found to be well converged with respect to the system size as well as the MC steps.

The multiband Hubbard Hamiltonian considered for the microscopic understanding of the system is given as, 
\begin{align*}
  H  &= \sum_{i,l,\sigma} \epsilon^{f}_d d^{\dagger}_{il\sigma} d_{il\sigma} + \sum_{i,l,\sigma} \epsilon^{m}_d d^{\dagger}_{il\sigma} d_{il\sigma} +	
     \sum_{i,l,\sigma} \epsilon_p p^{\dagger}_{il\sigma} p_{il\sigma} + \nonumber
     \sum_{i,l,\sigma} \epsilon_s s^{\dagger}_{il\sigma} s_{il\sigma}\\ \nonumber 
     &- \sum_{i,j,l_{1},l_{2},\sigma} \left( t^{l_{1}l_{2}}_{i,j,pd_f} d^{\dagger}_{il_{1}\sigma} p_{jl_{2}\sigma} + {\bf H.c.} \right) 
   - \sum_{i,j,l_{1},l_{2},\sigma} \left( t^{l_{1}l_{2}}_{i,j,pd_m} d^{\dagger}_{il_{1}\sigma} p_{jl_{2}\sigma} + {\bf H.c.} \right) \\ \nonumber
     &- \sum_{i,j,l_{1},l_{2},\sigma} \left( t^{l_{1}l_{2}}_{i,j,pp} p^{\dagger}_{il_{1}\sigma} p_{jl_{2}\sigma} + {\bf H.c.} \right) 
   - \sum_{i,j,l_{1},l_{2},\sigma} \left( t^{l_{1}l_{2}}_{i,j,sd_f} d^{\dagger}_{il_{1}\sigma} s_{jl_{2}\sigma} + {\bf H.c.} \right) \\ \nonumber
   &- \sum_{i,j,l_{1},l_{2},\sigma} \left( t^{l_{1}l_{2}}_{i,j,sd_m} d^{\dagger}_{il_{1}\sigma} s_{jl_{2}\sigma} + {\bf H.c.} \right) \\ \nonumber
   &+ \sum_{\alpha \beta \gamma \delta,\sigma_{1}\sigma_{2}\sigma_{3}\sigma_{4}} U^{\alpha \beta \gamma \delta}_{dd,f}   d^{\dagger}_{\alpha \sigma_1} 
       d^{\dagger}_{\beta \sigma_2} d_{\gamma \sigma_3} d_{\delta \sigma_4} \\ \nonumber
   &    + \sum_{\alpha \beta \gamma \delta,\sigma_{1}\sigma_{2}\sigma_{3}\sigma_{4}} U^{\alpha \beta \gamma \delta}_{dd,m}   d^{\dagger}_{\alpha \sigma_1} 
       d^{\dagger}_{\beta \sigma_2} d_{\gamma \sigma_3} d_{\delta \sigma_4} \\ 
\end{align*}

where $d^{\dagger}_{il\sigma} \left(d_{il\sigma} \right)$ creates (annihilates) an electron with spin $\sigma$ in the $l$th $d$ orbital on transition metal 
site in the $i$th unit cell while $p^{\dagger}_{il\sigma} \left( p_{il\sigma} \right)$/$s^{\dagger}_{il\sigma} \left(s_{il\sigma} \right)$ 
creates (annihilates) an electron with spin $\sigma$ in the $l$th $p$/$s$ orbital on oxygen atom in the $i$th unit cell. Here the index $f$ and $m$ represent the Fe and Re 
sites respectively.
The hopping interaction strengths ($t$) were kept fixed at the values obtained from fitting the nonmagnetic $ab$ $initio$ band structure of Sr$_2$FeReO$_6$ within a tight binding 
model having Fe $d$, Re $d$ and O $s$ and $p$ states in the basis. These hopping interactions were parametrized in terms of Slater-Koster parameters\cite{ch_sfro_ref38} dp$\sigma$, 
dp$\pi$, pp$\sigma$, pp$\pi$ and ds$\sigma$. 
The onsite energies were however varied, though the same values were taken for both spin channels. The Coulomb matrix elements were parameterized in  terms of the Slater-Condon integrals\cite{ch_sfro_ref25} $F^0$, $F^2$ and $F^4$ exploiting
its rotationally invariant nature and there values were determined from atomic Hartree-Fock calculations. 
As the value of $F^0$ is decreased significantly from its atomic value due to screening, 
it was varied and used to determine the multiplet averaged Coulomb 
interaction  $U$. As the Hund's exchange interaction strength is related to the Slater Condon integrals by the 
relation $J_h =  \frac{F^2 + F^4}{14}$, a reduction factor was used to get the desired exchange splitting observed in the $ab-initio$ spin 
polarized calculations. For Fe $d$ orbitals,a value of  $J_h$ of 0.8 eV was considered while at the Re site it is varied from 0.24 eV to 0.4 eV. 

The four fermion Coulomb interaction terms in the above Hamiltonian
were reduced to quadratic terms assuming a mean field approximation~\cite{ch_sfro_ref39, ch_sfro_ref40, prb4d5d}.
The expectation values of the order parameters entering the mean field decoupled Hamiltonian were determined 
self-consistently till an energy convergence of 10$^{-5}$ eV was achieved. Total energy and density of states were calculated using the tetrahedron method of 
integration for a 10 $\times$ 10 $\times$ 10 k-mesh.

\section{Results and Discussion}
 
An ordered double perovskite structure consists of two different transition metal atoms B and B$'$ arranged alternately, each with a local environment involving BO$_6$ and B$'$O$_6$ octahedra. The crystal structure of Sr$_2$FeReO$_6$ is simple cubic with space group Fm-3m~\cite{sfro_exp}. We have optimised both the unit cell parameters
as well as the internal positions within {\it ab-initio} electronic structure calculations with $U$(Fe) = 2.5 eV.  
The theoretically optimized structural information are given in Table~\ref{chpsfro_table.0}. 

\begin{table*}[!ht]\centering
\begin{tabular}{ cccc  }
\hline \hline
Structural details & Experimental & Experimental & Theoretical \\
\hline
Space group & Fm-3m &  I 4/m &Fm-3m \\
   &   &  & \\
a = b (\AA{}) & 7.888 (= c) & 5.561 & 7.890  (= c)\\ 
c  (\AA{}) & -- & 7.901 & -- \\ 
   &   &   & \\
$\alpha$ = $\beta$ = $\gamma$ (\textdegree) & 90 & 90 & 90\\
   &   &   \\
Bond lengths (\AA{}) &  &    & \\
Re-O &  1.70  & 1.959/1.962 &1.94   \\
Fe-O & 2.24   & 1.985/1.991 &  2.01   \\
   &   &  & \\
Fe-O-Mo Bond angles (in degrees) &   &   &\\
  In-plane &  180  & 170 & 180  \\
   Out-of-plane & 180 & 180 & 180\\
   &   &   &\\
Cell volume (\AA{}$^3$)  & 490.8  & 488.8 & 491.17 \\
\hline \hline
\end{tabular}
\caption{Experimental and optimized/theoretical structural details for Sr$_2$FeReO$_6$.  In second and third column, experimental details given for cubic ~\cite{sfro_exp} and tetragonal ~\cite{sfro_tetra} structures respectively.
Fe-O and Re-O bond lengths are quite similar for the experimental tetragonal phase and our optimized crystal structure of Sr$_2$FeReO$_6$.
phase}
\label{chpsfro_table.0}
\end{table*}

 The experimental unit cell parameters are also given for comparison. As one can see, in SFRO all the Fe-O-Re bond lengths are 180\textdegree{} while all the Fe-O or Re-O bondlengths are identical within the respective 
octahedra as is evident from Table~\ref{chpsfro_table.0}. Further there was no substantial change in the experimentally observed lattice constants found after structural optimization. However, there is substantial change in the Fe-O and Re-O bond lengths after the optimization. We use this optimized structure in the rest of our calculations. Obtained Fe-O and Re-O bond lengths in for our optimized structure 
is quite similar  with the ones found in tetragonal phase of Sr$_2$FeReO$_6$. This similarity among the bond lengths of the tetragonal crystal structure along with the fact that the tetragonal distortion is 
quite small in magnitude ($\sim$ 0.011 \AA{}) justify use of optimized crystal structure in our calculations. 
To understand the high magnetic ordering temperature of this system, the magnon dispersions were calculated along various symmetry directions in the 
Brillouin zone of SFRO (Fig. 1). Each point in Fig. 1 corresponds to a different
magnetic configuration identified by {$\vec q$} and has been calculated for a $U$ of 2.5 eV on the Fe $d$ states. Clearly, {$\vec q$}=0, ($\Gamma$) which represents the ferromagnetic state has the lowest energy.

\begin{figure}[!htbp]
\centering
\includegraphics[height=5.5cm,width=8.0cm]{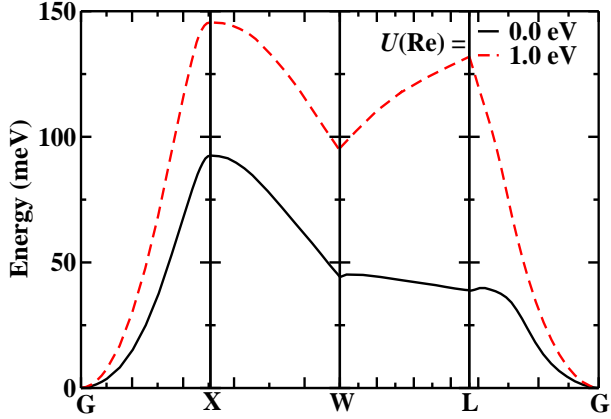}
\caption{Magnon dispersion plot for Sr$_2$FeReO$_6$ for $U$(Fe) = 2.5 eV and $U$(Re) = 0 and 1 eV. 
The G-point energy is the ferromagnetic ground state energy which gets higher stability with application of $U$ = 1 eV on Re site.}
  \label{fig.1}
\end{figure}

 This is in agreement with the 
experimental findings. The value of magnetic moment at Fe and Re sites are 3.9 $\mu_B$ and -0.95 $\mu_B$ respectively. 
The induced moment on the Re site is opposite to
the Fe moments and  originates from Fe-Re hopping as described in Ref\cite{ch_sfro_ref16}. This is evident from the 
density of states calculated for a perfectly antiferromagnetic configuration. In that case one had no exchange splitting of the Re-$d$ states, which one would otherwise find if there was a 
component emerging from an intrinsic exchange splitting.

\begin{figure}[ht]
\centering
\includegraphics[height=6.5cm,width=9.0cm]{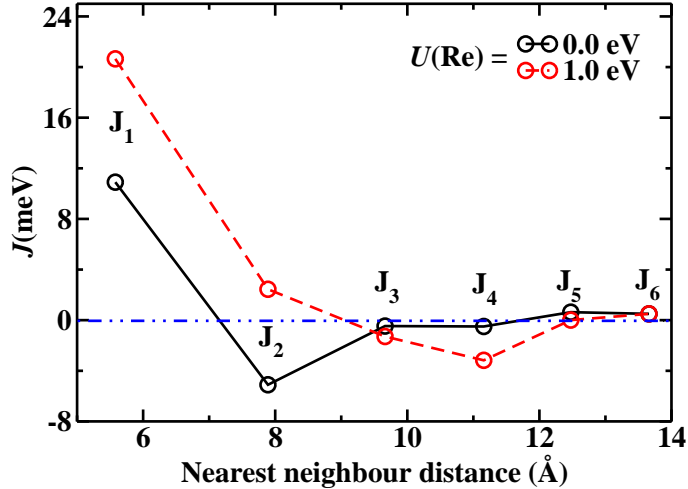}
\caption{ Fe-Fe interatomic exchange interaction parameters ($J$) of Sr$_2$FeReO$_6$ as a function Fe-Fe nearest neighbor distance (in \AA{}).}
  \label{fig.2}
\end{figure}

\begin{figure}[ht]
\centering
\includegraphics[height=8.0cm, width=10.0cm]{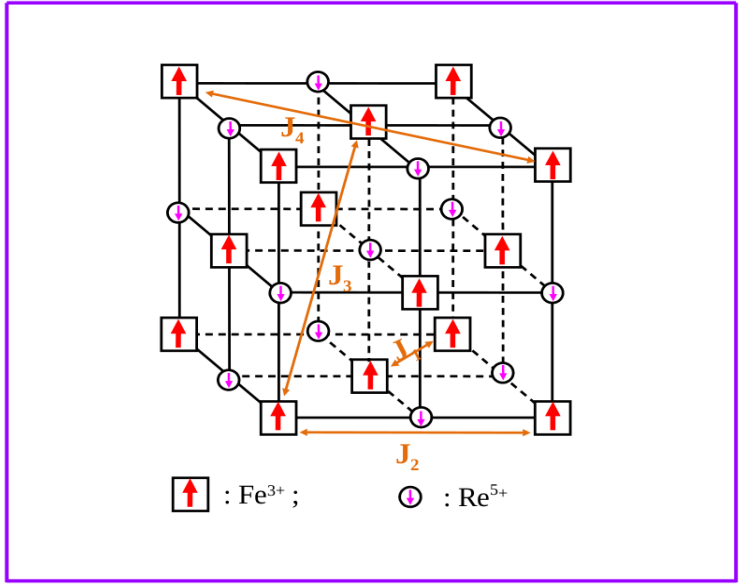}
\caption{A schematic to show the first four Fe-Fe interatomic exchange interaction strengths ($J$'s) in the simple cubic lattice of 
Sr$_2$FeReO$_6$.}
\label{neighbors}
\end{figure}

The real space exchange interaction strengths ($J$'s) are then estimated from an inverse Fourier 
transformation.  A plot of the different $J$'s as a function of distance is given in Fig.~2. 
We found that in this case when there is no $U$ on the Re-$d$ states, $J_1$ which is the 
most dominant one was found to be 10.9 meV and is ferromagnetic while $J_2$ is -5.1 meV and is antiferromagnetic in nature. All other J's are negligibly small. 
A schematic showing the first four exchange interactions and the pathways they represent is shown in Fig~\ref{neighbors}.
One should mention here that the separation between the Fe atoms connected via $J_1$, $J_2$, $J_3$ 
and $J_4$ are $\sim$ 5.58 \AA{}, 7.89 \AA{}, 9.66 \AA{} and 11.16 \AA{} respectively. Using these values for the exchange 
interaction strengths in a Monte Carlo calculation, one finds a magnetic transition temperature of 285~K. 
As shown by Brey {\it et al.}, application of $U$ on the nonmagnetic Mo site will penalize double occupancy. 
It consequently leads to the ferromagnetic state being favored. Taking a clue from this hypothesis, we applied a $U$ of 1 eV on the Re-$d$ states while a $U$ on Fe-$d$ states was kept 
fixed at the value of 2.5 eV and recalculated the magnon spectra shown in Fig~\ref{fig.1} (red dashed lines). A plot of estimated values of J's in this case is shown in 
Fig~\ref{fig.2}. 

\begin{figure}[ht]
\centering
\includegraphics[height=6.0cm,width=8.0cm]{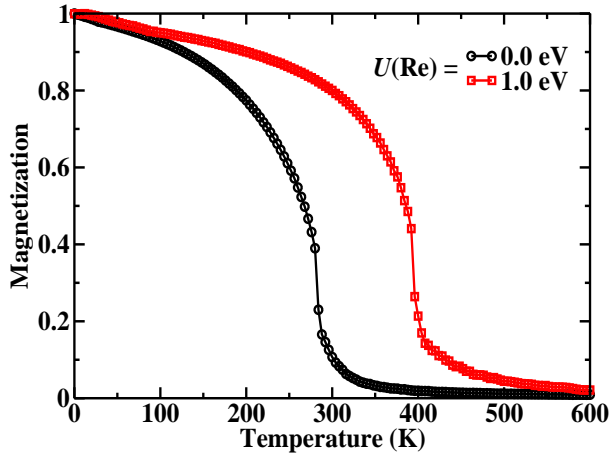}
\caption{ Variation of normalized magnetization as a function of temperature calculated for $U$(Fe) = 2.5 eV and $U$(Re) = 0 and 1 eV using Monte Carlo simulations for Sr$_2$FeReO$_6$.}
\label{fig.3}
\end{figure}

A clear enhancement in the stability of the ferromagnetic configuration can be found in this case when compared to the previous case of no $U$ on Re. 
The effect of $U$ on the Re site is also manifested in the exchange parameters. 
An analysis of the estimated $J$'s in this case yields mainly two important points. First is the enhancement in the magnitude of the most dominant nearest neighbor ferromagnetic 
exchange interaction strength $J_1$. Its value $\sim$ 21 meV is nearly doubled with a $U$(Re) = 1 eV from its value in the absence of $U$ on Re. Most interestingly, $J_2$ which was 
antiferromagnetic in the previous case is now found to be ferromagnetic in the present case. The estimated T$_c$ with this new set of exchange interaction strengths is 395 K 
which is in reasonable agreement with the experimental value of 401 K. The magnetic moment on the Re site is enhanced to -1.04 $\mu_B$ in this case while it is the same for the 
Fe atoms when there was no $U$ on the Re-d states. A plot of the normalized magnetic moment variation as a function of temperature in both the cases when there is no $U$ on 
Re and when a $U$ of 1 eV is considered on the Re-$d$ states is shown in Fig~\ref{fig.3}. There is indeed a clear enhancement of the T$_c$.

\begin{figure}[ht]
\centering
\includegraphics[height=5.0cm,width=10.0cm]{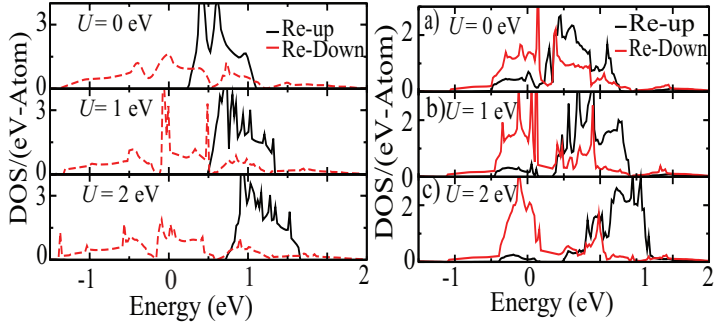}
\caption{Atom projected Re-$d$ density of states (DOS) plots from $ab$ $initio$ calculations for Sr$_2$FeReO$_6$. The plot on the left is for a ferromagnetic configuration of Fe spins while the plot in the right is from  the closest competing antiferromagnetic state of Fe spins. The zero of energy is the Fermi energy.}
  \label{fig.4}
\end{figure}

To explain the role of a $U$ on the Re $d$ states, we went to examine the Fe and Re atom projected density of states in both the cases with and without a $U$ on Re-$d$ 
states for the ferromagnetic ground states, a plot of which is shown in left panel of Fig.~\ref{fig.4}. We also include
the case where a $U$ of 2 eV was considered on the Re $d$ states.
The large exchange splitting at the Fe site and an induced small exchange splitting at Re site is evident from the plot.

As we can see in the up spin channel, Fe states are fully occupied verifying the 
3$d^5$ ionic configuration of Fe while the Re states are unoccupied and are $\sim$ 0.6 eV above the Fermi level. With the application of $U$ on Re, the unoccupied Re-up spin states 
are pushed up away from the Fermi level. A $U$ of 1 eV on Re pushes the Re $d$ states by $\sim$ 0.25 eV from its position for the case when there was no $U$. 
In the down spin channel, there is hardly any significant change in Fe and Re-$d$ density of states with the variation of $U$ on the Re 
site. They are found to be pinned to their positions in the absence of $U$. Thus, the small shifts in the Re-up spin density of states above the Fermi energy as 
function of $U$ on the Re $d$ states alone is therefore insufficient to describe the substantial gain in the stability of the ferromagnetic ground state found by examining the magnon 
dispersion shown in Fig.~\ref{fig.1}. For example, the gain is $\sim$ 50 meV at the X and W points while it is even more than $\sim$ 85 meV at the L-point over the $U$ = 0 values. 
We went on to examine the density of states of the closest competing antiferromagnetic state.
The spins on the Re site are antiferromagnetically coupled to the spins on the Fe sites. 
We consider antiferromagnetic configurations in which we have flipped the spin of two out of the six Fe nearest neighbors of a Re
atom and  made them ferromagnetic. As this is a collinear antiferromagnetic configuration with the smallest number of spins around Re flipped, one would expect this 
to be the closest lying antiferromagnetic excited states.

The atom projected density of states for Fe and Re atoms are then calculated by varying the $U$ on the Re site. The calculated density of states 
for the Fe-$d$ states have almost 
similar features as found for the ferromagnetic configuration in Fig.~\ref{fig.4}. Similar to the ferromagnetic case, these states 
are unaffected by the application of $U$ on Re atom and hence are not shown here. It is the Re-$d$ density of states shown in the right panel of Fig.~\ref{fig.4}, 
which have the most dramatic effects 
in this antiferromagnetic configuration with application of $U$ on it. 
When there is no $U$ on the Re states, both spin channels are found to be contributing at the Fermi level as shown in the 
right panel (a) of Fig.~\ref{fig.4}.
The induced exchange splitting at the Re site, which was present in the ferromagnetic case has reduced substantially. This provides pathways for both, the ferromagnetically 
coupled as well as the antiferromagnetically coupled spins. This explains the case of reduced magnetic transition temperature. 
The Re-up spin channel gets substantially depleted at the Fermi level with the application of $U$ on Re-$d$ states as shown in the right panel (b) of Fig.~\ref{fig.4}. The majority 
weight of the Re-up spin states is now much above the Fermi level bringing an effective exchange splitting at the Re site. Because of this the energy gain from delocalization 
of antiferromagnetically coupled Fe spins is significantly reduced. The exchange splitting at the Re site, further enhances the energy gain for ferromagnetically coupled spins.
This is manifested in terms of enhanced $J_1$ and a ferromagnetic $J_2$ in this case and consequently a higher T$_c$. This picture gets even more 
clear by looking at Re density of states with a $U$ of 2 eV at its $d$ states where the effects are more pronounced.
A comparison of the stability of the ferromagnetic states with respect to this antiferromagnetic state as a function of $U$(Re) is tabulated in Table~\ref{chpsfro_table.1}. 

\begin{table*}[ht]\centering
\begin{tabular}{@{}ccccccccc@{}} \toprule
\multicolumn{3}{c}{AFM-FM (eV)} \\
\hline
U$_{Re}$ (eV) & Sr$_2$FeReO$_6$ & Sr$_2$FeMoO$_6$ \\
\colrule
0.0 & 0.085 & 0.129 \\
1.0 & 0.138 & 0.143 \\
2.0 & 0.175 & 0.158 \\

\botrule
\end{tabular}
\caption{Change in stability of the ferromagnetic state calculated with respect to closest competing antiferromagnetic state from $ab$ $initio$ calculations as function of $U$ on the Re/Mo-$d$ states for 
Sr$_2$FeReO$_6$ and Sr$_2$FeMoO$_6$ double perovskites.}
\label{chpsfro_table.1}
\end{table*}

\begin{figure}[ht]
\centering
\includegraphics[height=5.5cm,width=7.5cm]{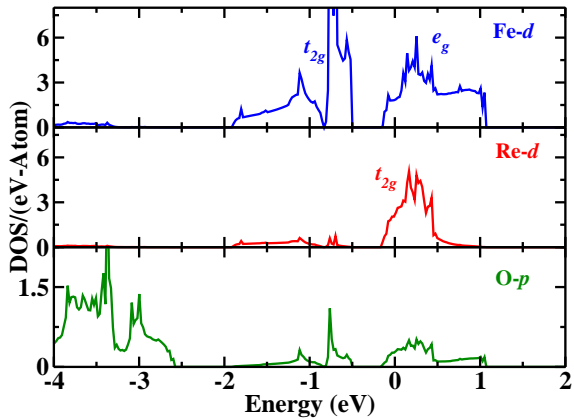}
\caption{Fe-$d$, Re-$d$ and oxygen-$p$  projected partial density of states for non-magnetic Sr$_2$FeReO$_6$ from  $ab$ $initio$ calculations. The zero of energy is the Fermi energy.}
  \label{fig.5}
\end{figure}

In order to explicitly understand the role of the exchange splitting at the nonmagnetic Re site, we further went on to develop a simple model which can define this system correctly. 
For this purpose, we start with calculating the $ab$ $initio$ atom projected density of states within a non-magnetic calculation for Sr$_2$FeReO$_6$ and is given in Fig. \ref{fig.5}.
The Fe $d$ states are 
found in the energy window from -2 eV to 1 eV. The octahedral field of the six oxygens which are the nearest neighbors of Fe lift the degeneracy of the $d$ orbitals and one has 
the $t_{2g}$ orbitals contributing in the energy window from -2 to -0.5 eV. The $e_g$ states contribute in the energy window from $\sim$ 0 to 1 eV. 
We find a similar degeneracy 
lifting of the $d$ states on Re, with the $t_{2g}$ states contributing in the energy window from $\sim$ 0 to 0.5 eV. 
Examining the O $p$ states, one finds that their primary 
contribution is below $\sim$ -3 eV and extends to -8 eV below the Fermi level.
There is a small admixture of the O $p$ states in the energy window contributed primarily 
by the Fe and Re $d$ states. This arises from $p$-$d$ hybridization and one finds that the contributions are larger for the Fe $e_g$ states as well as the Re $t_{2g}$ states 
which have the larger hopping matrix elements associated with them. Surprisingly we find the energy window over which the Fe $t_{2g}$ states contribute comparable with that 
in which Fe $e_g$ states contribute. This seems surprising and could emerge from the smaller energy denominator for the delocalization of an electron in  the Fe $t_{2g}$ states 
via the Re $t_{2g}$ states than the Fe $e_g$ states via the levels with the same symmetry on Re. 

\begin{figure}[ht]
\centering
\includegraphics[height=5.5cm,width=7.5cm]{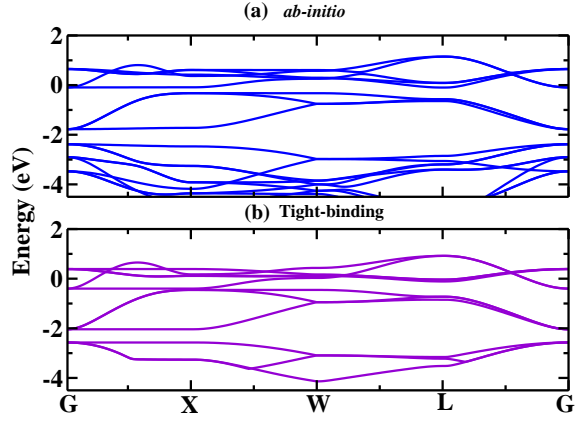}
\caption{ Non-magnetic band dispersions from (a) $ab$ $initio$ , (b) tight binding calculations with Fe-$d$, Re-$d$ and O-$s$ and $p$ basis states  
along various symmetry directions in Brillouin zone of Sr$_2$FeReO$_6$. The zero of energy is the Fermi energy.}
  \label{fig.6}
\end{figure}

 The {\it ab initio} band dispersions along various symmetry directions for nonmagnetic Sr$_2$FeReO$_6$ are shown in Fig.~\ref{fig.6} 
in the energy window -4 to 2 eV. As Fe $d$, Re $d$ and O $p$ states were found to contribute in this energy window, we included these states in the basis of tight
  binding model considered. 
We also included the $s$ states on the oxygen as discussed in the methodology section. 
The parameters entering the tight 
binding Hamiltonian should reproduce the band structure from the $ab$ $initio$ calculations. A least square error minimization procedure was used. The best fit
parameters are given in 
Table \ref{chpsfro_table.2}. 

\begin{table*}[!ht]\centering
\setlength\tabcolsep{4pt}
\begin{tabular}{@{}rrrrrcrrrcrrrcrrr@{}}\toprule
& \multicolumn{3}{c}{Onsite energies}  & \phantom{abc} & & \multicolumn{3}{c}{Fe-O} & \phantom{abc}& \multicolumn{3}{c}{Re-O} &
\phantom{abc} & \multicolumn{3}{c}{O-O} \\
 \cline{2-5} \cline{7-9} \cline{11-13}  \cline{15-17}
& \multicolumn{2}{c}{Fe-$d$} & Re-$d$ & O-$p$ && $dp\sigma$ & $dp\pi$ & $ds\sigma$ && $dp\sigma$ & $dp\pi$ & $ds\sigma$ && $pp\sigma$ & $pp\pi$ & \\  
\colrule 
& $t_{2g}$ & $e_g$  & $t_{2g}$& & \\ \cline{1-4} 
& 2.46 & 3.34 & 3.90 & 0.68 &&  1.90 & -1.495 & -1.44 && 3.05 & -1.28 & -0.76 && 0.75 & -0.164 \\
\botrule
\end{tabular}
\caption{Onsite energies and various hopping interaction strengths (in eV) parameterized in terms of Slater-Koster parameters obtained by fitting $ab$ $initio$ band structure 
of non-magnetic Sr$_2$FeReO$_6$ employing  
a tight binding model with Fe, Re -$d$ and O $s$ and $p$ states in the basis.}
\label{chpsfro_table.2}
\end{table*}

The $dp\pi$ interactions between the Fe $d$ and O $p$ states are significantly enhanced compared to a ratio of 0.5 for $dp\sigma/dp\pi$ interactions 
that one usually expects explaining partially the similar bandwidths that we find for Fe $t_{2g}$ and Fe $e_g$ states.
$ds\sigma$ interactions control the crystal field splitting between the $t_{2g}$ and $e_g$ orbitals and are found to be considerable for the Fe $d$ states also, compared to 
the Re $d$ states as the Re $e_g$ states were not included in the fitting. Examining the onsite energies, one finds that one has the Fe $t_{2g}$ states, Fe $e_g$ states and 
then the Re $t_{2g}$ states as expected. 

\begin{table*}[!ht]\centering
 \setlength \tabcolsep{4pt}
\begin{tabular}{@{}rrrrcrrrcrrrcrrr@{}}\toprule
&  \multicolumn{3}{c}{Onsite energies} & \phantom{abc} & \multicolumn{3}{c}{Fe-Fe} & \phantom{abc}& \multicolumn{3}{c}{Fe-Re} &
\phantom{abc} & \multicolumn{3}{c}{Re-Re}\\
 \cline{2-4} \cline{6-8} \cline{10-12} \cline{14-16}
& \multicolumn{2}{c}{Fe-$d$} & Re-$d$ && $dd\sigma$ & $dd\pi$ & $dd\delta$ && $dd\sigma$ & $dd\pi$ & $dd\delta$ && $dd\sigma$ & $dd\pi$ & $dd\delta$ \\ 
\colrule 
& $t_{2g}$ & $e_g$ & $t_{2g}$& &  \\ \cline{2-4}
& 4.45 & 6.39 & 5.10  && -0.05 & 0.05 & -0.004 && -0.866 & 0.298 & 0.0 && -0.199 & 0.008 & -0.02 \\

\botrule
\end{tabular}
\caption{Onsite energies and various hopping interaction strengths (in eV) parameterized in terms of Slater-Koster parameters obtained by fitting $ab$ $initio$ band structure 
of non-magnetic Sr$_2$FeReO$_6$ employing 
a tight binding model with only Fe and Re -$d$ states in the basis.}
\label{chpsfro_table.3}
\end{table*} 

\begin{figure}[!htbp]
\centering
\includegraphics[height=5.0cm,width=7.5cm]{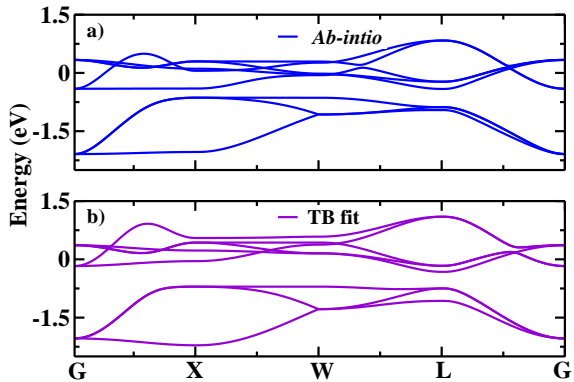}
\caption{Non-magnetic band structure from (a) $ab$ $initio$ , (b) tight binding calculations with only Fe and Re $d$ basis states, plotted
along various symmetry directions in Brillouin zone of Sr$_2$FeReO$_6$. The zero of energy is the Fermi energy.}
  \label{fig.7}
\end{figure}

The charge 
transfer energy between the O $p$ states and Fe $t_{2g}$ states is small but positive. In order to examine the role of explicitly including the O $p$ states in the basis, 
especially when the charge transfer energy is so small, a mapping was carried out onto a $d$-only model. The parameters found to give the best fit of the $ab$ $initio$ 
band structure shown in Fig \ref{fig.7}(a) are given in Table \ref{chpsfro_table.3}. The tight binding band structure calculated for the best fit parameters are given in 
Fig. \ref{fig.7}(b).

\begin{figure}[ht]
\centering
\includegraphics[height=5.5cm,width=8.0cm]{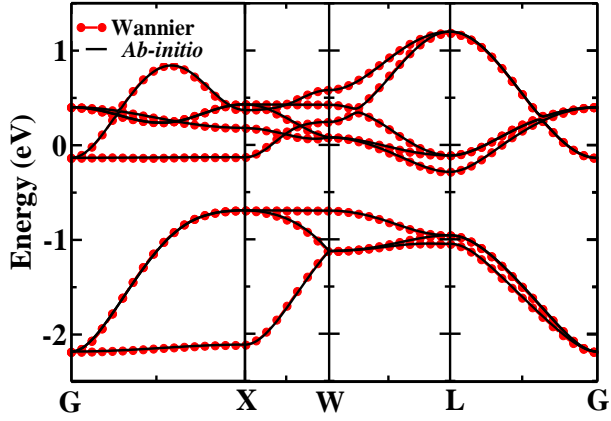}
\caption{ Non-magnetic $ab$ $initio$ band structure fitted with a Wannier functions based tight binding model considering Fe and Re -$d$ orbitals in the basis.  The zero of energy is the Fermi energy.}
  \label{fig.8}
\end{figure}

As is evident, we 
include explicitly the Fe $d$ and Re $d$ states while constructing the tight binding Hamiltonian with $d$-$d$ interactions. Interestingly we find that the Re $t_{2g}$ levels 
lie in between the Fe $t_{2g}$ and Fe $e_g$ levels. This is because in the effective $d$-only model, we are approximating the positions of the $d$ levels to their positions 
after their interaction with oxygen. The matrix elements for Fe $e_g$ and O $p$ interactions are larger, resulting in larger shifts compared to the interaction between 
Re $t_{2g}$ and O $p$ states. This leads to negative energy difference between the Fe $e_g$ and the Re $t_{2g}$ states. 
In order to verify if the placing of the Re $t_{2g}$ levels below the Fe $e_g$ levels is not an artefact of the fitting procedure, a similar analysis was carried out for a 
$d$ only mapping using maximally localized wannier functions\cite{ch_sfro_ref44} for the radial parts of the basis functions. A comparison of the $ab$ $initio$ band structure and the 
fitted band structure are given in Fig.~\ref{fig.8}. 

Here we find that the fit is reasonable which gives us confidence in examining the onsite energies. We find that the difference between 
the Re $t_{2g}$ and the Fe $e_g$ levels is negative, $\sim$ -0.7 eV in agreement with what we had earlier. 
In order to determine the appropriate model, a multiband Hubbard 
model is considered with the same value of $U$ and the electronic structure for a given set of parameters is calculated using a mean field decoupling scheme for the four Fermion 
terms. The hopping interaction strengths are kept fixed at the values obtained by fitting the nonmagnetic band structure.

\begin{figure}[ht]
\centering
\includegraphics[height=9cm,width=12.0cm]{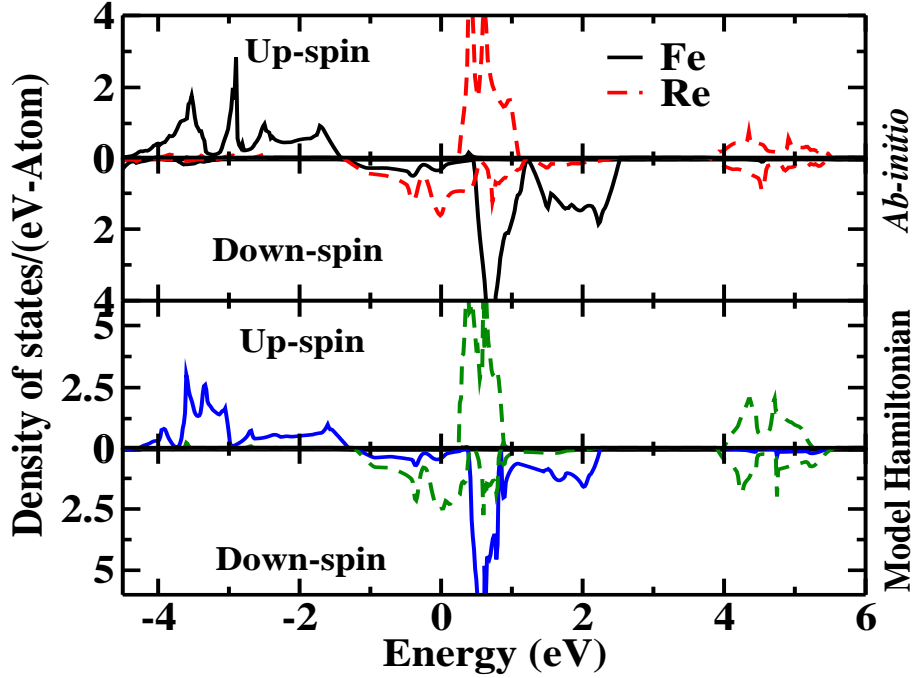}
\caption{ Atom projected density of states plots from $ab$ $initio$ calculations (upper two) and model Hamiltonian calculation (lower two) with Fe and Re $d$ 
states in the basis of Sr$_2$FeReO$_6$. The zero of energy is the Fermi energy.}
  \label{fig.9}
\end{figure}

 We consider the charge fluctuations at the Fe and Re 
sites as, 
\begin{eqnarray}
\Delta_{Fe \leftrightarrow O} = E_f(d^6 \bar{L}) - E_i(d^5 L^0) = \epsilon_d - \epsilon_p + 5U^{Fe} \\
\Delta_{Re \leftrightarrow O} = E_f(d^3 \bar{L}) - E_i(d^2 L^0) = \epsilon_d - \epsilon_p + 2U^{Re} 
\end{eqnarray}
Here $\epsilon_d$ and $\epsilon_p$ are the onsite energies of Fe/Re-$d$ and O-$p$ orbitals respectively. Hence $\Delta$ was kept fixed with $U_{Fe/Re}$ by changing the onsite energies 
in accordance to the above equations. 
The $J_h$ are varied till a good description of the $ab$ $initio$ density of states is obtained. A comparison is shown in Fig \ref{fig.9}. 

A $J_h$ of 0.8 eV was considered on Fe and this gave a good 
 description of the $ab$ $initio$ density of states. Additionally the charge transfer energy between both Fe and Re with oxygen were found to be 0.85 and 3.5 eV, both positive. 
 Additionally a $U$ of 0.26 eV and a small $J$ of 0.24 eV was used on the Re states to get the half-metallic character of the density of states. 
 As discussed previously with the $ab$ $initio$ results, a $U$ on the Re-$d$ states introduces an effective exchange splitting on the same, resulting in a gain for the 
 ferromagnetic ground state. We probe this fact independently using the model Hamiltonian that we have developed.
We calculate the stability of the ferromagnetic ground state with respect to the previously considered antiferromagnetic state with variation of 
 exchange splitting ($J_h$) at the Re site.

  The calculations were done with a fixed $U$ of 0.4 eV of the Re-$d$ states and keeping all the other parameters the same as their previous 
 values. 
 The obtained results are tabulated in Table~\ref{chpsfro_table.4}. 
 
 \begin{table*}[ht]\centering
\begin{tabular}{@{}ccccccccc@{}} \toprule

J$_h$ (eV) & AFM-FM (eV) \\
\colrule
0.24 & 0.134  \\
0.35 & 0.173  \\
0.50 & 0.186  \\

\botrule
\end{tabular}
\caption{Within a multiband Hubbard model, obtained change in stability of the ferromagnetic state with respect to closest competing antiferromagnetic state as function of 
exchange splitting at the Re site for Sr$_2$FeReO$_6$.}
\label{chpsfro_table.4}
\end{table*}

As we can see that with an increase in $J_h$ from 0.24 eV to 0.35 eV at the Re site, the stability of the ferromagnetic state is indeed enhanced by $\sim$ 40 meV. 
 This supports our previously given explanation for increase in the stability of ferromagnetic state with  application of $U$ on the Re site as it causes an effective 
 exchange splitting at the Re site.  This suggests a route to enhance the T$_c$ of these double perovskite systems by using magnetic atoms at the B$'$ site. 
 
  \begin{figure}[ht]
\centering
\includegraphics[height=5.0cm,width=10.0cm]{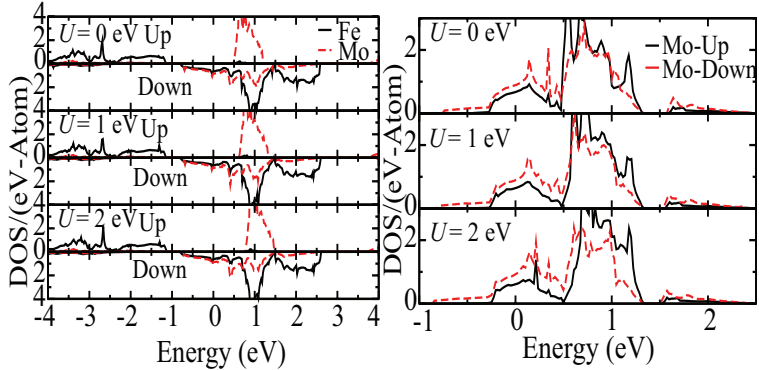}
\caption{Atom projected Fe, Mo-$d$ density of states plots from $ab$ $initio$ calculations for Sr$_2$FeMoO$_6$. The plot in the left is obtained for a 
 ferromagnetic configuration of Fe spins while the plot in the right is from  the closest competing antiferromagnetic state. The zero of energy is the Fermi energy.}
  \label{fig.10}
\end{figure}
 
Finally, a comparison with a similar system Sr$_2$FeMoO$_6$ (SFMO) has been drawn to explain the role two of electrons present in the Re-$d$ orbitals (Re$^{5+}$) in the case of SFRO. 
Change in the ferromagnetic stability with respect to the same antiferromagnetic configuration is calculated for SFMO and the results are tabulated in Table~\ref{chpsfro_table.1}. 
As one can see that in the case of SFMO, the stability of 
ferromagnetic state increases when a $U$ of 1 eV is applied at the Mo site, but the gain is much less than that of SFRO. While in the case of SFRO this gain was $\sim$ 50 meV, 
here it is only 14 meV. Further increase of $U$  at the Re site from 1 eV to 2 eV causes the stability to decrease in the case of SFRO while the stability is nearly constant in the 
case of SFMO. This weaker effect of $U$ on the Mo site in bringing an effective exchange splitting on it can easily be seen by analyzing the density of states for Fe, Mo-$d$ 
orbitals.  The features in the density of states shown in Fig.\ref{fig.10} for the ferromagnetic configuration (left panel) is almost similar to that of the SFRO with the 
only difference that the weight of Mo-$d$ orbitals in SFMO is less at the Fermi level when compared to Re-$d$ states in SFRO. This can be understood as the Re atom has two 
electrons in its $d$ orbitals in the ionic configuration of 5+ while Mo has only one electron in its 4$d$ orbital. In the antiferromagnetic configuration, $U$ of 1 eV on the 
Mo-$d$ has almost no effect on the its density of effects while a $U$ of 2 eV brings only a small change at the Fermi level explaining the less gain in stability of the 
ferromagnetic state with the variation of $U$.

\section{Conclusions}

In this study, using $ab$ $initio$ electronic structure calculations we have explained the role of electron-electron correlations at the nonmagnetic Re site in Sr$_2$FeReO$_6$ 
in the context of its 
high ferromagnetic ordering temperature. We show that  the role of $U$ on Re site is to introduce an effective exchange splitting reducing the gain from the antiferromagnetic 
pathways, providing higher stability to the ferromagnetic ground state. A simple multiband Hubbard like model is developed to independently verify this finding. Within this 
model, indeed, we find an increase in the stability of the ferromagnetic ground state with increase of exchange splitting at Re site. Role of two electrons of Re-$d$ orbitals 
result in substantial gain of the ferromagnetic stability and this is also contrasted with a similar compound, Sr$_2$FeMoO$_6$, which has single $d$ electron at the Mo site.

\end{document}